\documentclass[twocolumn,times]{aastex631}

\usepackage{xspace}
\usepackage{hyperref}
\usepackage{CJK}

\newcommand{\msun}{$M_{\odot}$}

\newcommand{\teff}{${T}_{\mathrm{eff}}$}
\newcommand{\logg}{$\log{g}$}

\newcommand{\tar}{SBSS\,1232+563\xspace}

\shorttitle{Transits of the white dwarf SBSS 1232+563}
\shortauthors{Hermes et al.}
\graphicspath{{./}{}}
\begin{document}
\begin{CJK}{UTF8}{gbsn}

\title{Sporadic Dips from Extended Debris Transiting the Metal-Rich White Dwarf SBSS 1232+563}

\correspondingauthor{J. J. Hermes}
\email{jjhermes@bu.edu}

\author[0000-0001-5941-2286]{J. J. Hermes}
\affiliation{Department of Astronomy, Boston University, 725 Commonwealth Ave., Boston, MA 02215, USA}

\author[0000-0001-9632-7347]{Joseph A. Guidry}\altaffiliation{NSF Graduate Research Fellow}
\affiliation{Department of Astronomy, Boston University, 725 Commonwealth Ave., Boston, MA 02215, USA}

\author[0000-0002-0853-3464]{Zachary P. Vanderbosch}
\affil{Division of Physics, Mathematics, and Astronomy, California Institute of Technology, Pasadena, CA 91125, USA}

\author[0000-0003-4903-567X]{Mariona Badenas-Agusti}
\affiliation{Institute of Astronomy, University of Cambridge, Madingley Road, Cambridge, CB3 0HA, UK}

\author[0000-0002-8808-4282]{Siyi Xu (许\CJKfamily{bsmi}偲\CJKfamily{gbsn}艺)} 
\affil{Gemini Observatory/NSF's NOIRLab, 670 N. A'ohoku Place, Hilo, Hawaii, 96720, USA}

\author[0000-0001-5745-3535]{Malia L. Kao}
\affiliation{Department of Astronomy, University of Texas at Austin, 2515 Speedway, Austin, TX 78712, USA}

\author[0000-0003-4189-9668]{Antonio~C.~Rodriguez}
\affil{Division of Physics, Mathematics, and Astronomy, California Institute of Technology, Pasadena, CA 91125, USA}

\author[0000-0002-1423-2174]{Keith Hawkins}
\affiliation{Department of Astronomy, University of Texas at Austin, 2515 Speedway, Austin, TX 78712, USA}

\begin{abstract}

We present the discovery of deep but sporadic transits in the flux of SBSS 1232+563, a metal-rich white dwarf polluted by disrupted exoplanetary debris. Nearly 25 years of photometry from multiple sky surveys reveal evidence of occasional dimming of the white dwarf, most notably evident in an 8-months-long event in 2023 that caused a $>$40\% drop in flux from the star. In-transit follow-up shows additional short-timescale (minutes- to hours-long) dimming events. TESS photometry suggests a coherent 14.842-hr signal that could represent the dominant orbital period of debris. Six low-resolution spectra collected at various transit depths over two decades show no evidence of significant changes in the observed elemental abundances. SBSS 1232+563 demonstrates that debris transits around white dwarfs can be sporadic, with many years of inactivity before large-amplitude dimming events.

\end{abstract}

% https://astrothesaurus.org/thesaurus/alphabetical-browse/

\section{Introduction} \label{sec:intro}

Although short-period exoplanets are expected to be engulfed as their host star evolves off the main sequence, distant-orbiting planets (e.g., beyond Mars in our Solar System) are expected to persist all the way to the white dwarf stage \citep{2008MNRAS.386..155S}. Surviving planets provide the gravitational impetus to scatter asteroids and planetesimals onto star-grazing orbits, where we observe the consequence of tidal shredding of these rocks by the central white dwarf. Disrupted material can form dusty disks detectable as an infrared excess, as well as gaseous disks with atomic lines detected in emission or absorption, and the material eventually works its way onto the white dwarf, polluting its surface; see reviews by \citet{2016RSOS....350571V} and \citet{2016NewAR..71....9F}.

The consensus picture of accretion of disrupted exoplanetary material onto the photosphere of metal-rich white dwarfs was corroborated by the discovery of deep, irregular transits from rocky debris towards the metal-rich white dwarf WD\,1145+017 recurring roughly every 4.5\,hr, close to the period expected for a circular orbit near the tidal disruption radius \citep{2015Natur.526..546V}. Several additional metal-rich white dwarfs that exhibit photometric transits have been discovered from transient surveys such as the Zwicky Transient Facility (ZTF), including both long-period transiting systems with dip recurrence timescales greater than 100\,d \citep{2020ApJ...897..171V,2021ApJ...912..125G}, as well as shorter-period systems with recurrence timescales less than 12\,hr \citep{2021ApJ...917...41V}.

Patterns in the debris transiting white dwarfs have so far been characterized by reproducibility over a few orbital cycles but wholesale changes over tens of orbital cycles, likely revealing the dynamics around these chaotic circumstellar environments. For example, drifting fragments from one 25-hr orbit to the next are seen in the transiting white dwarf WD\,1054-226 \citep{2022MNRAS.511.1647F}. Long-term changes are also frequently observed in these transiting systems. In fact, WD\,1145+017 has seen a drastic reduction in transit activity since roughly 2022 \citep{2024MNRAS.530..117A}.

One previously claimed candidate transiting debris system is the metal-rich white dwarf \tar. Not to be confused with the Sloan Digital Sky Survey (SDSS), \tar\ was first cataloged as a Second Byurakan Survey Star (SBSS), a program initiated in 1974 from the Byurakan Observatory in Armenia to search for galaxies, quasars, and other ultraviolet-bright objects \citep{2000A&AS..147..169B}. \tar\ was first spectroscopically classified as a DBAZ by \citet{2006ApJS..167...40E} from an early SDSS DR4 spectrum; the classification implies a helium-dominated (DB) spectrum with large contributions from hydrogen (A) and metals (Z\footnote{More recent references classify \tar\ as a DZBA given that metals are the dominant spectral component. Historically, this judgment has been done by visual inspection and is therefore somewhat subjective.}). An analysis of the SDSS spectrum along with the photometry plus the Gaia DR2 parallax by \citet{2019ApJ...885...74C} find \teff\ = $11790\pm420$\,K, \logg\ = $8.30\pm0.06$, log(H/He) = $-5.52\pm0.14$, and log(Ca/He) = $-7.41\pm0.06$, atmospheric parameters we adopt throughout this work.

Higher-resolution spectroscopy from Keck reveal eight heavy elements (O, Mn, Cr, Si, Fe, Mg, Ca, Ti) in addition to H and He in \tar, corresponding to a total accretion rate of material in excess of $5\times10^{9}$\,g\,s$^{-1}$ \citep{2019AJ....158..242X}. More than 60\% of the accreted material is O by mass, which implies that the majority of O delivered was not locked up in rocks such as MgO or SiO$_2$ \citep{2010ApJ...709..950K}. An oxygen excess can be interpreted as evidence of water-rich asteroid debris (e.g., \citealt{2013Sci...342..218F,2015MNRAS.450.2083R}). While the O excess in \tar\ appears robust \citep{2023MNRAS.519.2663B}, there is not enough H measured in the star for H$_2$O to be responsible for even 20\% of this O excess, making the source of excess O an open question \citep{2019AJ....158..242X}. %5e9 g/s implies >1 Myr to accrete the mass of the Martian moon Phobos.

The infrared photometry from WISE does not appear to follow the spectral energy distribution of a single white dwarf; the IR excess at the $W1$ (3.4\,$\mu$m) and $W2$ (4.5\,$\mu$m) bands was used by \citet{2011ApJS..197...38D} to argue that \tar\ is an excellent candidate to host a dusty debris disk. For a time it was claimed to be the most massive white dwarf with a disk \citep{2014ApJ...786...77B}, but that analysis used an incorrect mass determination from \citet{2006ApJS..167...40E}; the mass of \tar\ has been refined to a lower value of $0.77\pm0.04$\,\msun\ \citep{2019ApJ...885...74C}. Spitzer imaging confirms that the IR excess is real, arising from dusty debris \citep{2020MNRAS.496.5233S}. The Spitzer data also revealed 8\% (5$\sigma$) variability at 4.5$\mu$m, further evidence that the IR excess is from a debris disk (e.g., \citealt{2024ApJ...972..126G}).

Transits from planetary debris were first suspected in \tar\ based on its anomalously large photometric uncertainties in Gaia DR2 \citep{zenodo.4088554}. More detailed variability analysis also incorporating ZTF photometry made \tar\ a candidate for transiting debris, although more than 36\,hr of high-speed photometry from McDonald Observatory did not reveal any dramatic ($>$7\% depth) dimming events \citep{2021ApJ...912..125G}.

Interestingly, fits for the atmospheric parameters of \tar\ in \citet{2019ApJ...885...74C} were complicated by the variable photometry, who noted: ``The SDSS magnitudes for this object are $\sim$0.3 mag fainter than Pan-STARRS and Gaia, which leads to two different possible solutions. We could not reach a conclusion regarding this discrepancy...'' We propose here the solution: \tar\ undergoes sporadic and occasionally deep transits from orbiting debris.

In Section~\ref{sec:photometry} we detail the long-baseline survey photometry that shows sporadic dips from planetary debris transiting \tar. In Section~\ref{sec:highspeed} we detail more than 37.5\,hr of high-speed photometry collected to look for short-period transit events from the source, as well as six sectors of TESS observations. In Section~\ref{sec:spec} we put low-resolution spectroscopy collected on \tar\ over many years into the context of the observed transits. We conclude with a discussion in Section~\ref{sec:discuss}.

\section{Survey Photometry of SBSS 1232+563} \label{sec:photometry}

\begin{figure*}[t!]
    \centering
    {\includegraphics[width=0.995\textwidth]{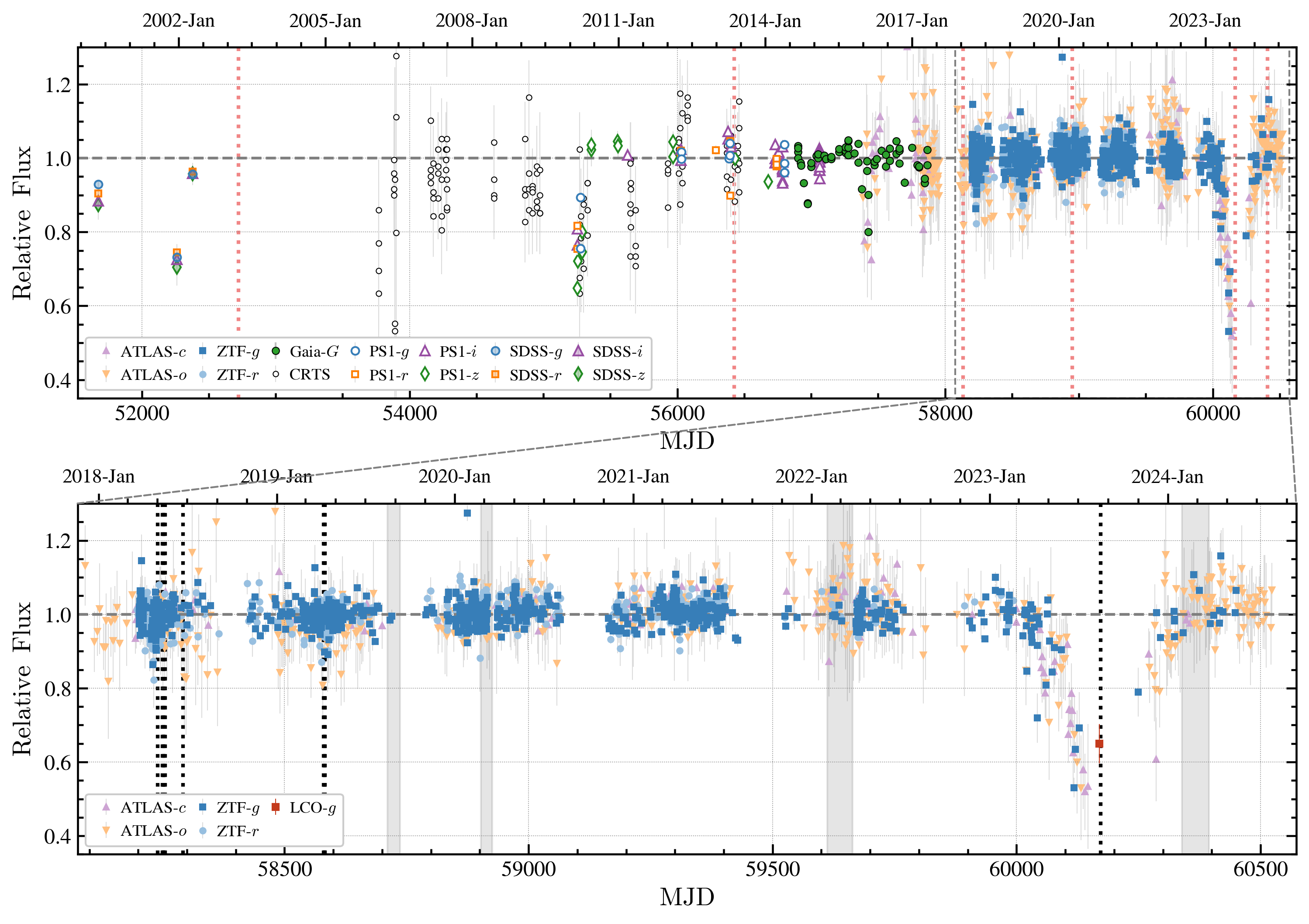}}
    \caption{The top panel shows survey photometry for \tar, detailed in Section~\ref{sec:photometry}, spanning 2000 to 2024 shows sporadic, deep dips. Low-resolution optical spectroscopy coincides with vertical dashed red lines, including a spectrum in 2023 collected in transit (see Section~\ref{sec:spec}). The bottom panel shows a zoom in from 2018 to mid-2024. The deepest dip began in roughly 2023~April and extended until roughly 2024~January (more than 8 months), assuming the object remained in transit while it was not visible behind the Sun. The observing windows of high-speed photometric follow-up from McDonald Observatory (detailed in Section~\ref{sec:mcd}) are marked with vertical dashed black lines. Observational coverage during six sectors with TESS (further discussed in Section~\ref{sec:tess}) is shown by gray shaded regions.}%
    \label{fig:allsurvey}
\end{figure*}

% ++++++++++++++++++ Median Filters ++++++++++++++++++ %
\begin{deluxetable}{lcccc}
\tablecaption{Median Survey Filter Photometry \label{tab:survey}}
\tabletypesize{\small}
\tablehead{
    \colhead{Survey}    & \colhead{Date Range}  &
    \colhead{Filter}   & \colhead{$\lambda_{\mathrm{cen}}$} & \colhead{Median}   \\ [-0.2cm]
    \colhead{}           & \colhead{}      &
    \colhead{}       & \colhead{(nm)}   &  \colhead{(mag)}           
}
\startdata
ZTF & $2018-2024$ & $g$ & 481 & 18.03 \\
    &             & $r$ & 644 & 18.20 \\
    &             & $i$ & 752 & 18.44 \\
ATLAS & $2015-2024$ & $c$ & 518 & 18.05 \\
      &             & $o$ & 663 & 18.28 \\
Gaia  & $2014-2017$ & $G$ & 673 & 18.04 \\
PS1   & $2010-2015$ & $g$ & 487 & 17.95 \\
      &             & $r$ & 622 & 18.14 \\
      &             & $i$ & 755 & 18.35 \\
      &             & $z$ & 868 & 18.55  \\
CRTS  & $2006-2013$ & $V$ & 550 & 18.01
\enddata
\tablecomments{CRTS photometry is unfiltered but the CCD response creates an effective $\lambda_{\mathrm{cen}}$ near the $V$-band \citep{2013ApJ...763...32D}}
\end{deluxetable}
% =================================================== %

Figure~\ref{fig:allsurvey} shows the full 25-year photometric baseline collected on \tar. We detail the median magnitudes in each filter used for the relative photometry in Table~\ref{tab:survey}. We further detail the survey photometry we have used, roughly in reverse chronological order:

\textbf{ZTF} The bulk of our photometry comes from ZTF partnership surveys \citep{2019PASP..131a8003M} which we extracted through forced PSF photometry (see the ZTF Forced Photometry Service (ZFPS) documentation\footnote{\url{https://irsa.ipac.caltech.edu/data/ZTF/docs/ztf_zfps_userguide.pdf}}) on the Gaia-measured centroid, yielding 1573 epochs from 2018~March through 2024~May across the ZTF $g$- and $r$- bands. We do not include the ZTF $i$-band light curve in our analysis due to a lack of epochs, most of which are concentrated years before the 2023 transit. We curate our photometry following the standard recommendations from the ZFPS documentation: we select only {\tt infobitssci}\,$=0$ measurements with S/N\,$>$\,5 and those brighter than the estimated limiting magnitude inflated by 0.5\,mag and do not perform other filtering on the light curves \citep[c.f.,][]{2021ApJ...912..125G}. We calibrate measured zero-point magnitudes from the difference imaging to absolute magnitudes and perform color corrections to the cataloged Pan-STARRS magnitudes. We finally convert these color-corrected magnitudes to relative fluxes by normalizing the measurements to the median ZTF magnitude and propagating the associated errors. We carried out this procedure for each other survey to do our analysis in relative flux space. ZTF epochs are not corrected for proper motion, sufficient for this 6-year baseline.

\textbf{ATLAS} Significant recent photometry comes from the Asteroid Terrestrial-impact Last Alert System (ATLAS), which is a system of four telescopes in Hawaii, Chile, and South Africa that scan the sky for moving objects, in the process producing large archives of photometry down to roughly $G<19$\,mag \citep{2018PASP..130f4505T}. We do forced photometry on the ATLAS reduced science images\footnote{\url{https://fallingstar-data.com/forcedphot/}} using the Gaia-measured centroid of \tar, taking into account the measured proper motion. We apply barycentric corrections to all timestamps due to the multi-site nature of ATLAS. Keeping measurements only with S/N\,$>$\,3, we bin the extracted cyan ($c$-band) and orange ($o$-band) photometry to within one day, respectively, to further boost signal. After binning, we convert all remaining magnitudes to fluxes using the median magnitude of the binned light curve. We retain 669 epochs from 2015~December through 2024~August.

\textbf{Gaia} For every object classified as a variable by a machine learning algorithm trained on constant objects, Gaia Data Release 3 (DR3) provided multi-epoch light curves \citep{2023AA...674A..13E}. Each individual epoch is observed through one of the Gaia $G$, $G_{{\rm RP}}$, or $G_{{\rm BP}}$ band-passes. More than 1300 high-confidence white dwarfs were flagged in Gaia DR3 as variables \citep{2024ApJ...967..166S}, including \tar\ (Gaia DR3 1571584539980588544). We accessed the epoch photometry of \tar\ via {\tt astroquery} \citep{2019AJ....157...98G}, yielding 58 $G$-band epochs sampled between 2014~September to 2017~April. We only assess the $G$-band photometry here, as it has more precision and the $G_{{\rm RP}}$ and $G_{{\rm BP}}$ span the same observing baseline. The median Gaia magnitude in Table~\ref{tab:survey} is reported in Vega magnitudes; all other surveys use the AB system.

\textbf{PS1} We acquired the Pan-STARRS (PS1) \citep{2016arXiv161205560C} DR2 light curve photometry via MAST. We imposed a quality check requiring photometry with ${\tt psfQfPerfect}>0.95$, yielding 73 points in $griz$, surveying dates between 2010~February to 2015~February. We plot on a relative flux scale the PSF-extracted magnitudes normalized to their mean AB magnitude for each bandpass \citep{2012ApJ...750...99T} reported in Table~\ref{tab:survey}.

%\begin{figure*}[t!]
%    \centering
%    {\includegraphics[width=0.995\textwidth]{sbss1232_lc_updated_zoom.png}}
%    \caption{Survey photometry for \tar\ from 2018 to mid-2024. The deepest dip began in roughly 2023~April and extended until roughly 2024~January (more than 8 months), assuming the object remained in transit while it was not visible behind the Sun. The observing windows of high-speed photometric follow-up from McDonald Observatory (detailed in Section~\ref{sec:mcd}) are marked with vertical dashed black lines. Observational coverage during six sectors with TESS (further discussed in Section~\ref{sec:tess}) is shown by gray shaded regions.}%
%    \label{fig:zoomsurvey}
%\end{figure*}

\textbf{CRTS} The Catalina Real-Time Transient Survey (CRTS) \citep{2009ApJ...696..870D} is a multi-telescope optical transient survey. We queried CRTS DR2 photometry through a cone search\footnote{\url{http://nunuku.caltech.edu/cgi-bin/getcssconedb_release_img.cgi}} on the J2000 coordinates of \tar, yielding 134 un-blended epochs in the unfiltered (effective $V$-band) from 2006~February through 2013~June. We only include the 20 epochs after 56100~MJD to calculate the median CRTS magnitude, as most of the previous epochs appear to be taken in transit.

\textbf{SDSS} The earliest photometry analyzed here was collected by SDSS on 2000~May, with two other epochs on 2001~December and 2002~April. We acquired $ugriz$ photometry at these epochs through an SQL query for {\tt thingID}\,$=512173411$. To put these SDSS photometry onto a consistent flux scale, we apply AB zero-point offsets ($u_{\text{AB}}=u_{\text{SDSS}}-0.040, i_{\text{AB}}=i_{\text{SDSS}}+0.015, z_{\text{AB}}=z_{\text{SDSS}}+0.030$, with no correction in $g$ and $r$, as defined in \citealt{2006ApJS..167...40E}) to the individual magnitude measurements before converting them to fluxes normalized to the respective Pan-STARRS mean magnitude (since it appears most SDSS photometry was taken in relatively deep transit).

\textbf{LCOGT} Seeing \tar\ enter transit, we proposed for additional monitoring with the Las Cumbres Observatory network (LCOGT). We obtained $3\times60$\,s exposures in the $g$-band on 2023 August 13 from the 1.0-meter at Tenerife North in 2.1\arcsec\ seeing, which confirmed \tar\ to be $g=18.51\pm0.06$\,mag, more than 30\% dimmer than the ZTF-$g$ median flux baseline.

\textbf{WISE} We inspected near-infrared photometry of \tar\ from the Wide-field Infrared Survey Explorer (WISE) at 3.4\,$\mu$m ($W1$-band) and 4.6\,$\mu$m ($W2$-band) from the NEOWISE mission \citep{2014ApJ...792...30M}. We queried the single-exposure image (hereafter referred to as L1b) photometry published within the 2024 NEOWISE data release from IRSA using {\tt unTimely} \citep{2023AJ....165...36M}. The unWISE co-adds from the 2024 NEOWISE data release have yet to be released, preventing us from reproducing the analysis from \citet{2024ApJ...972..126G}. We follow the methodology of \citet{2024ApJ...972..126G} and convert the queried $W1$ and $W2$ Vega magnitudes from the NEOWISE era (2013$-$2023) to fluxes in $\mu$Jy using the published conversions and subsequently sigma clip the photometry within a given epoch to 3$\sigma$ for each bandpass, which we bin into single measurements. We estimate errors on our photometry by taking the ratio of the standard deviation of the clipped photometry to the square root of the number of measurements at that epoch. We do not include the L1b photometry in Figure~\ref{fig:allsurvey} due to the large point-to-point scatter ($>$20\%). 

\section{High-Speed Photometric Follow-up} \label{sec:highspeed}

Initially motivated by the inflated Gaia DR2 photometric uncertainties \citep{zenodo.4088554}, we densely observed \tar\ over several nights in 2018~May and again in 2019~April, as detailed in Section~\ref{sec:mcd}, with epochs marked as dashed black lines in Figure~\ref{fig:allsurvey}. Additionally, the object is well-enough isolated that the Transiting Exoplanet Survey Satellite (TESS) produces useful photometry. Those observations are outlined in Section~\ref{sec:tess}.

\subsection{McDonald Observatory Time-Series Photometry} \label{sec:mcd}

%\vfill
% +++++++++++++++++++ Journal Obs +++++++++++++++++++ %
\begin{deluxetable}{lcccccc}[b!]
\tabletypesize{\small}
\tablecolumns{7}
\tablewidth{0.985\columnwidth}
\tablecaption{Journal of 2.1-m McD High-Speed Photometry. \label{tab:jour}}
\tablehead{
\colhead{UT Date} & \colhead{Duration} & \colhead{Seeing} & \colhead{Ap.} & \colhead{Exp.} & \colhead{P2P} & \colhead{$\Delta$Flux} \\ \colhead{} & \colhead{(hr)} & \colhead{(\arcsec)} & \colhead{(px)} & \colhead{(s)} & \colhead{(\%)} & \colhead{(\%)} }
\startdata
2018 May 2 & 1.76 & 1.4 & 3.5 & 20 & 2.3 & 6.2 \\
2018 May 12 & 3.42 & 1.3 & 4.0 & 20 & 1.2 & 6.9 \\
2018 May 14 & 4.37 & 1.6 & 4.5 & 20 & 1.7 & 6.2 \\
2018 May 15 & 4.62 & 1.5 & 4.0 & 30 & 1.2 & 5.5 \\
2018 May 17 & 4.01 & 1.0 & 3.5 & 10 & 1.1 & 3.2 \\
2018 Jun 23 & 2.37 & 1.1 & 3.0 & 20 & 1.3 & 5.0 \\
2019 Apr 7 & 2.50 & 1.7 & 5.0 & 30 & 1.3 & 5.3 \\
2019 Apr 9 & 7.13 & 1.3 & 3.5 & 30 & 1.0 & 3.3 \\
2019 Apr 10 & 6.01 & 1.4 & 3.5 & 30 & 1.2 & 5.4 \\
2023 Aug 16 & 1.36 & 1.2 & 3.5 & 5 & 2.6 & 20.8
\enddata
\tablecomments{Ap. is the radius in pixels of the circular aperture drawn for our photometry extraction. P2P is the average point-to-point scatter of each light curve binned to 1-minute cadence. $\Delta$Flux is the total flux excursion within the first 1\,hr of the 1-minute-binned light curve.}
\end{deluxetable}
% ==================================================== %

\begin{figure*}[t!]
    \centering
    {\includegraphics[width=0.995\textwidth]{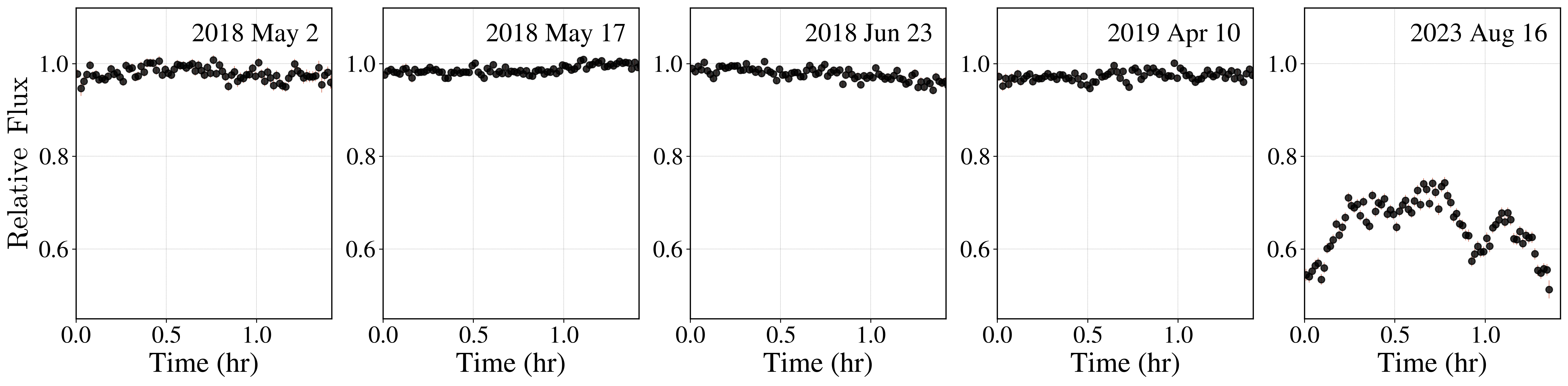}}
    \caption{The first 1.4 hr for five separate nights of high-speed photometry of \tar\ from the ProEM instrument on the 2.1-meter Otto Struve Telescope at McDonald Observatory. The left four panels were all collected out of transit and are representative for our five other epochs from 2018--2019, while the right panel was collected in 2023~August in deep transit and shows considerably more short-term variability. All data have been binned to 1-min and the closest measured nightly flux from Figure~\ref{fig:allsurvey} subtracted.}%
    \label{fig:mcd_lcs}
\end{figure*}

We collected high-speed follow-up time-series photometry of \tar\ for a total of more than 37.5 hours using the ProEM frame-transfer photometer mounted on the 2.1-m Otto Struve Telescope at McDonald Observatory. Our 10 individual epochs  are summarized in Table~\ref{tab:jour}. To reduce sky noise, we use a broad-bandpass, red-cutoff $BG40$ filter for all observations, which has a similar effective wavelength as the $g$ filter. As indicated by the vertical lines in Figure~\ref{fig:allsurvey}, all of these epochs in 2018 and 2019 were taken outside of a deep transit event; only our 2023~August observations were collected in transit. 

\citet{2021ApJ...912..125G} showed roughly 4\,hr from one of these McDonald runs from the night of 2019 April 9 in their Figure~6. We show five additional light curves in Figure~\ref{fig:mcd_lcs}. All data were collected in good conditions, with seeing generally $<$1.5\arcsec. We follow the reduction routine from \citet{2021ApJ...912..125G} by executing standard \textsc{iraf} routines to dark-subtract and flat-field, running \textsc{ccd\_hsp} \citep{2002A&A...389..896K} to do circular aperture photometry over a series of radii from $0.5-10.0$ pixels, and using {\tt phot2lc} \citep{phot2lc} to extract the light curve that minimizes the average point-to-point scatter.

We performed differential photometry on a stable, nearby comparison star. Some of our longer runs suffer from some differential extinction (there is an airmass-dependent color difference in our blue target and the available comparison stars). We usually divide out a second-order polynomial to account for these airmass changes over longer ($>$4\,hr) runs \citep{2009JPhCS.172a2081T}. Since we are interested in directly comparing possible short-period, transit-induced variability in our light curves, we do not divide out any low-order polynomial terms.

Figure~\ref{fig:mcd_lcs} shows five representative light curves, each to the same scale, showing the dramatic short-period, in-transit variability observed in 2023~August. We co-add all light curves into 1-minute bins to help directly compare each run. Despite suggestions of variability, we did not see evidence for major transits deeper than roughly 7\% in the 2018 and 2019 light curves of \tar.

Light curve metrics detailed in Table~\ref{tab:jour} measure the point-to-point (P2P) scatter of these 1-minute bins. We also measure the peak-to-peak variability for the first 1\,hr of each run ($\Delta$Flux), measuring the maximum flux excursion from the highest and lowest point of these 1-minute binned light curves. The McDonald/ProEM measurements reveal that the light curve collected in deep transit in 2023~August shows at least three times more extreme short-term (minutes-long) variability than any other epoch of observation in 2018 and 2019, all of which were collected out of major transit (see black dashed lines in Figure~\ref{fig:allsurvey}).

\subsection{Coherent Signals Present in TESS} \label{sec:tess}

Photometry from \tar\ was captured over six sectors by TESS \citep{2015JATIS...1a4003R}. Full-frame image (FFI) data were collected every 30-minutes during the first two sectors (Sectors 15 \& 22), observed during most of 2019~August and then most of 2020~May. The cadence of FFI images decreased to 10 minutes during observations in Sectors 48 \& 49, covering almost all of 2022~February to 2022~March. FFI image exposure times decreased again, to 200 seconds, for data collected in Sectors 75 \& 76 (observed from 2024~February to 2024~March).

In addition to FFI data, TESS targeted \tar\ (cataloged as TIC\,950506740, $T$=18.1\,mag) with 20-second cadence data. The white dwarf is incredibly faint for the effective 10-cm aperture of TESS, but it is in a relatively uncrowded field, which still makes the TESS photometry useful; the target contributes more than 35\% of the flux in the 5-pixel extraction aperture in Sector 75 as well as the 6-pixel aperture for Sector 76. There is only one ($G=20.6$) object in Gaia within 30\,arcsec of \tar, 21.5\arcsec\ to the SE. The nearest brighter stars are a $G=16.7$\,mag star 55.7\arcsec\ to the N and a $G=17.9$\,mag star 58.0\arcsec\ to the W.

\begin{figure}[t!]
    \centering
    {\includegraphics[width=0.995\columnwidth]{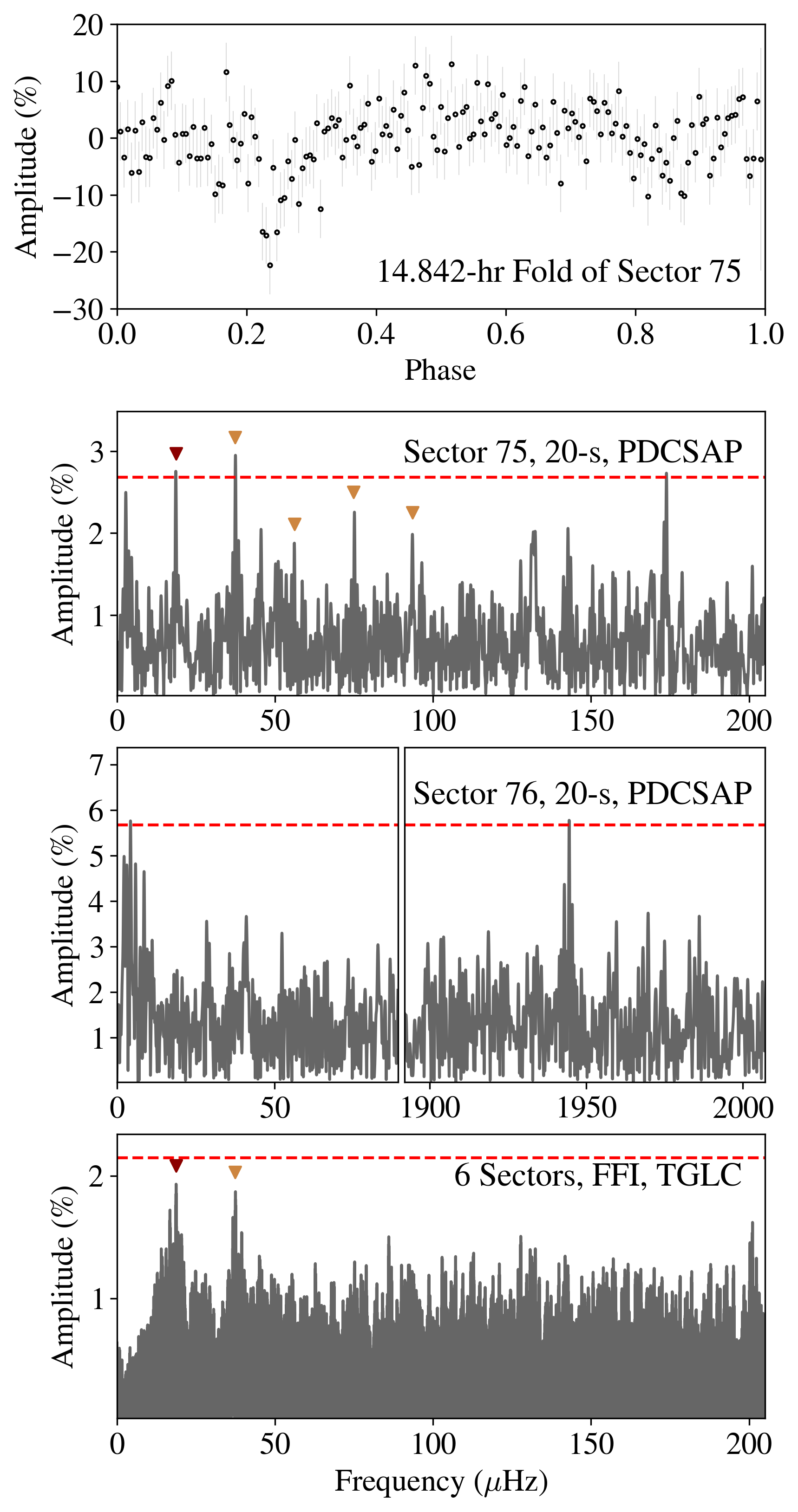}}
    \caption{The top panel shows 180 phase bins of the light curve of \tar\ from TESS Sector 75 folded at 14.842\,hr. The panels below show Lomb-Scargle periodograms of various TESS data. The second panel shows a periodogram of 20-s data from Sector~75, the highest-quality TESS data, where we detect significant signals at 14.842\,hr (18.7\,$\mu$Hz), the second harmonic at 7.421\,hr, and a (likely) non-harmonic signal at 95.8416\,min. We mark the fundamental with a maroon triangle and integer harmonics as orange triangles. The dashed red line is a bootstrapped 0.1\% False-Alarm Probability (above which peaks are formally significant). The third panel shows 20-s data from Sector 76. No significant peaks related to the 14.842\,hr fundamental are seen, though there is a significant signal at 1944.5 $\mu$Hz (8.57\,min) of uncertain origin. The bottom panel shows all FFI data. No formally significant signals are detected in earlier FFI data, but the 14.842-hr fundamental and its second harmonic are marginally detected in data from all six sectors.}%
    \label{fig:tess_ft}
\end{figure}

We extracted all six sectors of FFI data using the PSF fitting of background Gaia sources using the TESS-Gaia Light Curve (TGLC) software \citep{2023AJ....165...71H}. For our last two sectors we also analyzed the 20-second light curves from the Pre-search Data Conditioning Simple Aperture Photometry (PDCSAP) fluxes produced by the Science Processing Operations Center (SPOC) and hosted on MAST \citep{2016SPIE.9913E..3EJ}. Sectors 75 \& 76 suffer from significant scattered light, mostly from the Earth, yielding a poor duty cycle of 44.9\% and 38.8\%, respectively. 

We show in the middle panels of Figure~\ref{fig:tess_ft} Lomb-Scargle periodograms of the 20-second PDCSAP light curves. We also show empirical 0.1\% False-Alarm Probability for each sector, above which peaks have a $<$0.1\% chance to arise from noise, which were bootstrapped from $10{,}000$ periodograms of shuffling the fluxes but keeping the time sampling the same (e.g., \citealt{2019A&A...632A..42B}). 

Sector~75 reveals multiple significant signals: a first and second harmonic at $f_1=18.715$\,$\mu$Hz ($14.842$\,hr) and $2f_1=37.431$\,$\mu$Hz (see Table~\ref{tab:TESSperiods}). Peaks are also present at the 3rd, 4th, and 5th harmonics of the 14.842-hr fundamental signal, but they fall below the 0.1\% false-alarm threshold. These harmonics arise to reproduce the non-sinusoidal structure shown in the folded light curve in the top panel of Figure~\ref{fig:tess_ft}. Sector~75 also reveals a significant signal at $f_2=173.898$\,$\mu$Hz ($95.8416\pm0.0085$\,min) which does not appear harmonically related to the 14.842-hr fundamental. All significant peaks exceed 2.68\% amplitude (the false-alarm line). Sector~75 has the smaller point-to-point scatter and flux uncertainties than the other 20-s-cadence data in Sector~76, and was collected from 2024~January~30 to 2024~February~26, close to when the star returned to near-quiescent flux levels after the deep 2023 transit.

Data immediately thereafter in Sector 76 does not reveal significant peaks at any harmonics of the 14.842-hr fundamental signal. There is a formally significant peak at low-frequency (4.35\,$\mu$Hz) that, at 2.66\,d, is likely connected to background variability. More notable, there is a significant peak at $f_3=1944.521$\,$\mu$Hz ($8.57\pm0.14$\,min, see Table~\ref{tab:TESSperiods}). If harmonically related to the 14.842-hr fundamental, this would be the 103rd or 104th harmonic of $f_1$; although high harmonics of debris orbital frequencies have been observed in transiting debris systems \citep{2022MNRAS.511.1647F}, we merely suggest such a connection at this time.

We use \texttt{TESS\_localize} \citep{2023AJ....165..141H} to vet the origin of the significant signals in the 20-s TESS data. In Sector~75, we find that the first five harmonics of $f_1=18.715$\,$\mu$Hz have a relative likelihood of 1.000 ($p$-value = 0.044\footnote{The $p$-values from \texttt{TESS\_localize} estimate the fraction of fit locations that would have been less likely than the actual fit location under the hypothesis that each source in the field is the variable source, such that higher $p$-values are more reliable \citep{2023AJ....165..141H}.}) of arising from \tar. In another independent test, we find the signal $f_2=173.898$\,$\mu$Hz has a relative likelihood of 1.000 ($p$-value = 0.055) of arising from \tar. In Sector~76, we find the $f_3=1944.521$\,$\mu$Hz signal is confidently associated with \tar, with a relative likelihood of 1.000 ($p$-value = 0.962). We have also performed periodograms on the background flux values of our PDCSAP light curves and do not find peaks at $f_1-f_3$ or their harmonics.

% +++++++++++++++++++ TESS Periods +++++++++++++++++++ %
\begin{deluxetable}{lccc}[t!]
\tabletypesize{\small}
\tablecolumns{4}
\tablewidth{0.985\columnwidth}
\tablecaption{Significant signals in TESS photometry. \label{tab:TESSperiods}}
\tablehead{\colhead{ID} & \colhead{Period} & \colhead{Frequency} & \colhead{Amplitude}  
\\ & \colhead{(hr)} & \colhead{($\mu$Hz)} & \colhead{(\%)} }
\startdata
\multicolumn{4}{c}{\bf TESS 20-s Cadence, Sector 75} \\
%\hline\\
$f_1$ & 14.842 $\pm$ 0.035 & 18.715 $\pm$ 0.044 & 2.56 $\pm$ 0.49 \\
$2f_1$ & 7.4211 $\pm$ 0.0084 & 37.431 $\pm$ 0.042 & 2.67 $\pm$ 0.49 \\
$f_2$ & 1.59736 $\pm$ 0.00014 & 173.898 $\pm$ 0.015 & 2.84 $\pm$ 0.19 \\
%\hline\\
\multicolumn{4}{c}{\bf TESS 20-s Cadence, Sector 76} \\
%\hline\\
$f_3$ & 0.1428515 $\pm$ 0.0000030 & 1944.521 $\pm$ 0.041 & 5.8 $\pm$ 1.0
\enddata
\end{deluxetable}
% ==================================================== %

If persistent, such high-amplitude signals should show up in our other datasets. However, we do not identify periodic signals in either the ZTF $g$, $r$, and combined $g$+$r$ photometry, nor in the ATLAS $c$, $o$, and combined $c$+$o$ photometry. Only aliases from the diurnal survey pattern appear. We similarly do not identify signals at 14.842-hr, 95.8416-min, or 8.57-min in the periodograms of either of the combined 2018~May and 2019~April McDonald ProEM light curves. We do, however, find that the combined McDonald 2018~May light curve has a best period of 8.616\,hr, found from a period search using the Stellingwerf phase-dispersion minimization \citep{1978ApJ...224..953S} period-finding algorithm as implemented in \texttt{astrobase} \citep{wbhatti_astrobase}. This trial period is a marginally better fit than the second- and third-best periods of 6.456\,hr and 7.512\,hr, respectively. It is possible that the 7.512-hr signal is an alias of the half harmonic of the 14.842-hr signal seen in TESS, but none of these periods are formally significant.

That we cannot see the high-amplitude signals at 14.842\,hr, 95.8416\,min, and 8.57\,min observed by TESS in other datasets challenges that these are intrinsic astrophysical variability toward \tar. However, the Sector 75 \& 76 TESS data are some of the only photometry collected very near if not in the transit event that began in 2023~April. It is possible these signals are transient and only manifest in transit. Our 1.4-hr of photometry in 2023~August from McDonald Observatory taken in deep transit are too short to confirm any periodic signals.

\begin{figure*}[t!]
    \centering
    {\includegraphics[width=0.995\textwidth]{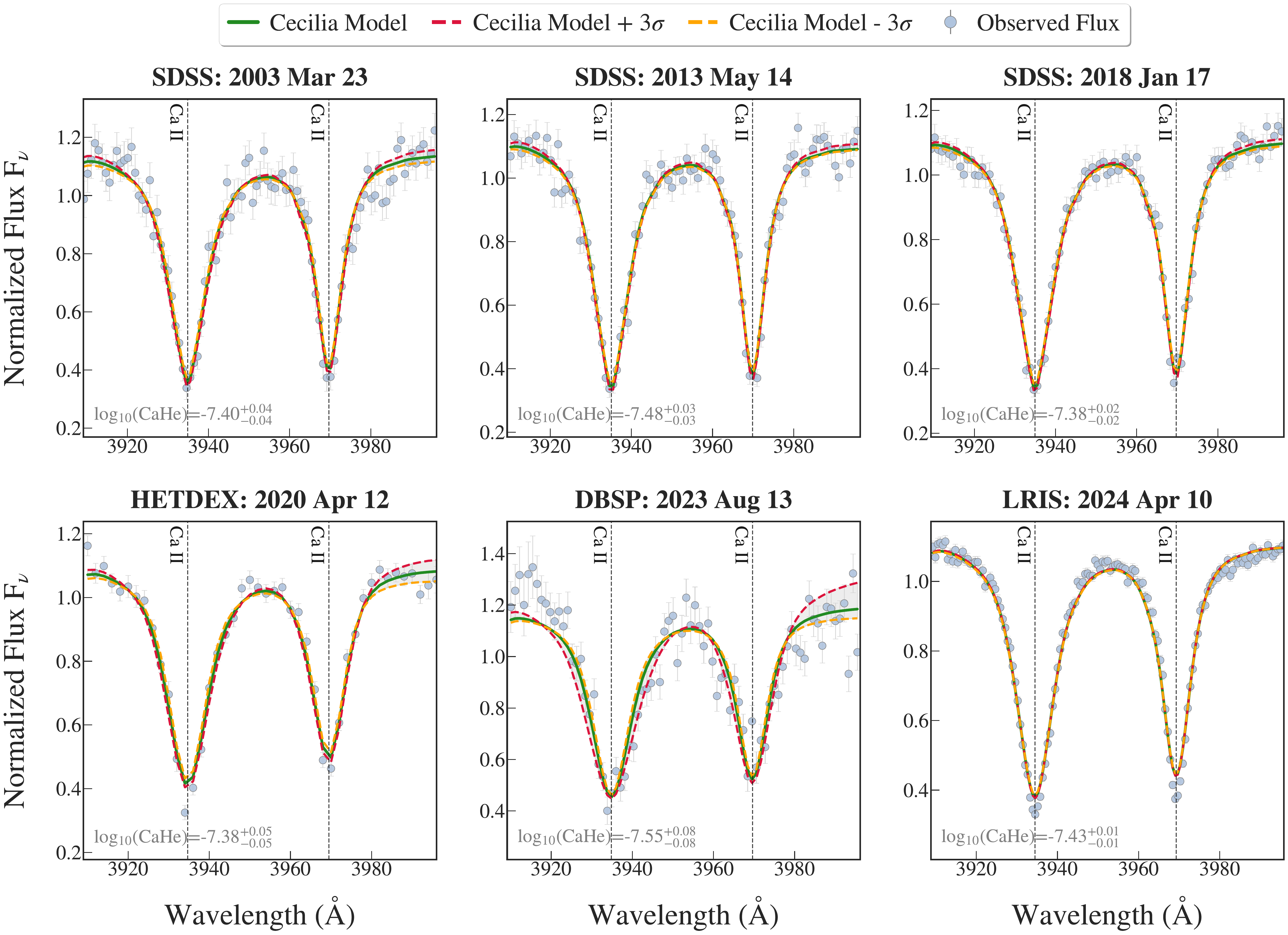}}
    \caption{Six epochs of low-resolution spectroscopy collected on \tar, at epochs denoted in Figure~\ref{fig:allsurvey}. We zoom in here on the Calcium H \& K lines, with the best-fit \texttt{cecilia} \citep{2024MNRAS.529.1688B} model overplotted in green. The values reported for each epoch include the statistical MCMC uncertainties; full abundances including systematic uncertainties for all detected metals are detailed in Table~\ref{tab:abund}. }%
    \label{fig:cecilia}
\end{figure*}

\section{Search for Spectroscopic Changes} \label{sec:spec}

There is considerable spectroscopy of \tar, some archival and spanning more than two decades and taken both in and out of transit, marked as six separate dashed gray lines in Figure~\ref{fig:allsurvey}. Each of these spectra, focused around the Calcium H \& K lines, are shown in Figure~\ref{fig:cecilia}. Basic properties of the spectra are outlined in Table~\ref{tab:abund}.

% ++++++++++++++++++ Abundances ++++++++++++++++++ %
\begin{deluxetable*}{lccccccc}
\tabletypesize{\footnotesize}
\tablecolumns{9}
\tablewidth{0.995\textwidth}
\tablecaption{Spectroscopic information and abundance measurements of \tar\ at different epochs.} \label{tab:abund}
\tablehead{
\colhead{Value} & \colhead{APO (SDSS)} & \colhead{APO (BOSS)} & \colhead{APO (BOSS)} & \colhead{McD (HETDEX)} & \colhead{Palomar (DBSP)} & \colhead{Keck (LRIS)} & \colhead{\citet{2019AJ....158..242X}} }
\startdata
Observation Date & 2003 Mar 23 & 2013 May 14 & 2018 Jan 17 & 2020 Apr 12 & 2023 Aug 13 & 2024 Apr 10 & $2015-2017$ \\
Wavelengths (\AA) & $3800-9200$ & $3610-10{,}140$ & $3610-10{,}140$ & $3500-5500$ & $3200-10{,}800$ & $3150-10{,}270$ & $3200-10{,}000$ \\
Effective Resolution & 1800 & 2000 & 2000 & 750 & 1000 & 1100 & $40{,}000$ \\
Signal-to-Noise & 34.4 & 48.6 & 51.8 & 34.9 & 17.9 & 158.4 & 68.0 \\
%Signal-to-Noise Per Pixel & 21.9 & 29.7 & 31.7 & 30.2 & 6.8 & 62.4 \\
Fraction of Flux & $\sim$95\% (w/in 1\,yr) & 99\% (w/in 1\,d) & 99\% (w/in 1\,d) & 100\% (w/in 1\,d) & 65\% (w/in 1\,d) & 97\% (w/in 1\,d) & $94-100$\% \\
$\log_{10}\rm{(Ca/He)}$ & $-7.40\pm0.13$ & $-7.48\pm0.13$ & $-7.38\pm0.12$ & $-7.38\pm0.13$ & $-7.55\pm0.15$ & $-7.43\pm0.12$ & $-7.69\pm0.05$ \\
$\log_{10}\rm{(Mg/He)}$ & $-6.01\pm0.12$ & $-6.07\pm0.12$ & $-6.03\pm0.11$ & $-6.06\pm0.12$ & $<-6.09$ & $-6.02\pm0.11$ & $-6.09\pm0.05$ \\
$\log_{10}\rm{(Fe/He)}$ & $-6.37\pm0.15$ & $-6.27\pm0.13$ & $-6.54\pm0.14$ & $-6.37^{+0.16}_{-0.15}$ & $<-5.73$ & $-6.21\pm0.12$ & $-6.45\pm0.11$ \\
$\log_{10}\rm{(O/He)}$ & $<-4.80$ & $-5.13^{+0.18}_{-0.17}$ & $-5.02^{+0.18}_{-0.17}$ & $<-3.99$ & $<-4.40$ & $-5.32\pm0.14$ & $-5.14\pm0.15$ \\
$\log_{10}\rm{(Si/He)}$ & $-6.28^{+0.16}_{-0.17}$ & $-6.23\pm0.14$ & $-6.40^{+0.15}_{-0.16}$ & $<-6.01$ & $<-5.63$ & $-6.32\pm0.13$ & $-6.36\pm0.13$ \\
$\log_{10}\rm{(Ti/He)}$ & $<-8.02$ & $<-8.61$ & $<-8.47$ & $<-8.23$ & $<-7.72$ & $-8.86\pm0.13$ & $-8.96\pm0.11$ \\
$\log_{10}\rm{(H/He)}$ & $<-5.41$ & $-5.61\pm0.13$ & $-5.82^{+0.16}_{-0.18}$ & $<-4.97$ & $<-5.30$ & $-5.63\pm0.11$ & $-5.90\pm0.15$
\enddata
\tablecomments{All analyses made with \texttt{cecilia} \citep{2024MNRAS.529.1688B} except for those from Keck HIRES/ESI data by \citet{2019AJ....158..242X}, who perform a line-by-line $\chi^2$ approach that applies the same weights to each spectral feature as described in \citet{2012ApJ...749....6D}. The Signal-to-Noise is measured per resolution element in the range from $5200-5300$\,\AA. Fraction of flux shows the measured relative flux from Figure~\ref{fig:allsurvey} and when that measurement was made compared to the spectroscopy.}
\end{deluxetable*}
% ================================================= %

\textbf{SDSS} The oldest spectrum of \tar\ was collected on 2003~March~23 and released as part of SDSS DR4 \citep{2006ApJS..167...40E}, collected with the original SDSS spectrograph. SDSS revisited the white dwarf two more times, on 2013~May~14 and 2018~January~17; the latter two spectra were collected with the BOSS spectrograph \citep{2012AJ....144..144B}. All SDSS spectral data were obtained via SciServer \citep{2020A&C....3300412T}. 

\textbf{HETDEX} \tar\ was observed on 2020 April 12\footnote{Almost no observatories besides HET were operating in April 2020 due to the COVID-19 pandemic.} by the Hobby-Eberly Telescope Dark Energy Experiment (HETDEX, \citealt{2008ASPC..399..115H}). The white dwarf was included in the second data release of the HETDEX Survey, which performed a re-extraction of HETDEX stellar spectra to account for Gaia proper motions \citep{2021ApJ...911..108H}.

\textbf{DBSP} When \tar\ entered transit in 2023 we triggered spectroscopy with the Double Spectrograph (DBSP) on the 200-inch at Palomar Observatory \citep{1982PASP...94..586O}. Our two 15-min exposures were collected using a 1.5\arcsec\ slit on 2023 August 13 using the 316-line grating blazed at 7500\,\AA\ for the red arm and the 600-line grism blazed at 4000\,\AA\ on the blue arm; due to a CCD defect on the red arm, there is a gap in coverage from $5550-5730$\,\AA. We reduced and co-added the spectra using \texttt{DBSP-DRP}\footnote{\url{https://dbsp-drp.readthedocs.io/en/stable/index.html}}, which is built upon the \texttt{PypeIt} package \citep{2020JOSS....5.2308P}. This was our only available spectroscopic observation of \tar\ in deep transit.

\textbf{LRIS} Our final spectrum in this work was collected using the Low-Resolution Imaging Spectrometer (LRIS) mounted on the Keck-I 10-meter telescope on Mauna Kea \citep{1995PASP..107..375O,2010SPIE.7735E..0RR}. We collected two 10-min exposures on 2024~April~10 using a 1.0\arcsec\ slit with the D560 dichroic. We used the 600 line~mm$^{-1}$ grism blazed at 4000\,\AA\ on the blue arm and the 400 line~mm$^{-1}$ grating blazed at 8500\,\AA\ on the red arm. We reduced and co-added the spectra using the LRIS automated reduction pipeline \citep[LPipe,][]{2019PASP..131h4503P} using the flux standard G191-B2B.

We analyze the abundances (and possible abundance changes) of the low-resolution spectra of \tar\ using \texttt{cecilia}, a machine-learning-based pipeline for measuring the elemental abundances of He-rich white dwarfs \citep{2024MNRAS.529.1688B}. In fact, the most recent SDSS spectrum of \tar\ collected in 2018~January was used as a test-case in Section~5.2 of \citet{2024MNRAS.529.1688B}. We extend that analysis to all low-resolution spectra in Figure~\ref{fig:cecilia}.

Table~\ref{tab:abund} summarizes our MCMC best-fit parameters using \texttt{cecilia}. We follow identically the steps outlined in Section~5.2 of \citet{2024MNRAS.529.1688B}, fixing the atmospheric parameters (\teff\ = $11{,}787$\,K, \logg\ = 8.30) based on those of \citet{2019ApJ...885...74C}, although we caution that these parameters, especially the \logg, likely have extra uncertainty due to the variable photometry of the system and the most accurate mass and temperature will require a more detailed analysis. In summary, before each fit, we slightly processed all spectra by removing regions around known skylines and tellurics. We then executed individual MCMC runs with 50 walkers, 3000 links, and a 20\% burn-in, imposing wide chondritic priors on elements that are typically easy to detect (Ca, Mg, Fe, O, Si; prior width of 2 dex), and narrower priors on the remaining elements (Ti, Cr, Mn, Ni: 0.5 dex). This configuration was sufficient to achieve convergence and took between 1.8\,hr (HETDEX) and 6.2\,hr (LRIS) to complete.

The uncertainties given in Table~\ref{tab:abund} reflect adding the statistical MCMC uncertainties ($\sigma_{\rm{stat, MCMC}}$), systematic  uncertainties from \texttt{cecilia}'s machine learning-based interpolation  ($\sigma_{\rm{sys,ML}}$, as detailed in Table~5 of \citealt{2024MNRAS.529.1688B}), and systematic atmospheric model uncertainties ($\sigma_{\rm{sys, model}}=0.10$\,dex) in quadrature. We assume that an element is detected when its total uncertainty is less than 0.20 dex and confirm this detectability threshold through a visual inspection of the spectra. In Table~\ref{tab:abund}, we report 3$\sigma$ upper limits on any elements with total uncertainties above 0.20 dex. Our best 3$\sigma$ upper limits for Cr, Mn, or Ni come from our LRIS spectrum: $\log_{10}\rm{(Cr/He)} < -7.98$, $\log_{10}\rm{(Mn/He)} < -8.15$, and $\log_{10}\rm{(Ni/He)} < -6.92$.

When detected, the measured elemental abundances all agree within 1$\sigma$ at all epochs, and are always within 1$\sigma$ consistent with the abundances determined from the highest-resolution spectra of \tar\ collected by Keck/HIRES presented in \citet{2019AJ....158..242X}, which we also include in Table~\ref{tab:abund}. The abundance measurements by \texttt{cecilia} for Ca at all six epochs agree within 1$\sigma$ with the value of $\log_{10}\rm{(Ca/He)} = -7.41\pm0.06$ from the \citet{2019ApJ...885...74C} analysis of the 2013 SDSS spectrum\footnote{We note that all but one Ca measurement is $>$1$\sigma$ more abundant than Keck/HIRES spectra from 2015 April 11 and 2016 April 1 found by \citet{2019AJ....158..242X} to have $\log_{10}\rm{(Ca/He)} = -7.69\pm0.05$ determined from an independent analysis not using \texttt{cecilia}. This discrepancy does not show up for any other elements.}.

The six low-resolution spectra analyzed here show remarkable consistency in the observed elemental abundances (especially Ca, Mg, Fe, O, and Si). All elements except H have long ($>$0.3\,Myr) settling times \citep{1986ApJS...61..177P}. It is worth noting that the highest-resolution (R\,=\,$40{,}000$) spectra from \citet{2019AJ....158..242X} show no extra broadening or other evidence of circumstellar absorption lines.

\section{Discussion \& Conclusions} \label{sec:discuss}

We have detected sporadic, extended transits in multicolor optical photometry of the metal-rich white dwarf \tar, most likely caused by occultations due to disrupted rocky debris. Dense observations from the transient surveys ZTF and ATLAS show that for more than 5 years (from 2018~January to 2023~January) the white dwarf was within a few percent of constant flux. \tar\ then entered a roughly 8-months-long, nearly 50\% deep transit for most of 2023. The long duration of the event suggests an extended cloud likely required to follow a highly eccentric orbit with an extreme apastron. % (tens of au). 

High-speed photometry from McDonald Observatory shows that there is higher-amplitude, short-term (minutes to hours) variability in-transit (in 2023) compared to data taken out of transit (in 2018 and 2019; Figure~\ref{fig:mcd_lcs}). ZTF\,J0923+4236, a white dwarf that shows sporadic transits from planetary debris at weeks-long intervals, similarly shows heightened short-timescale variability in-transit, relative to its near-flat out-of-transit flux baseline \citep{2024MNRAS.530..117A}. This suggests that the transit events are an inhomogeneous cloud of debris unevenly blocking the white dwarf.

\begin{figure*}[t!]
    \centering
    {\includegraphics[width=0.995\textwidth]{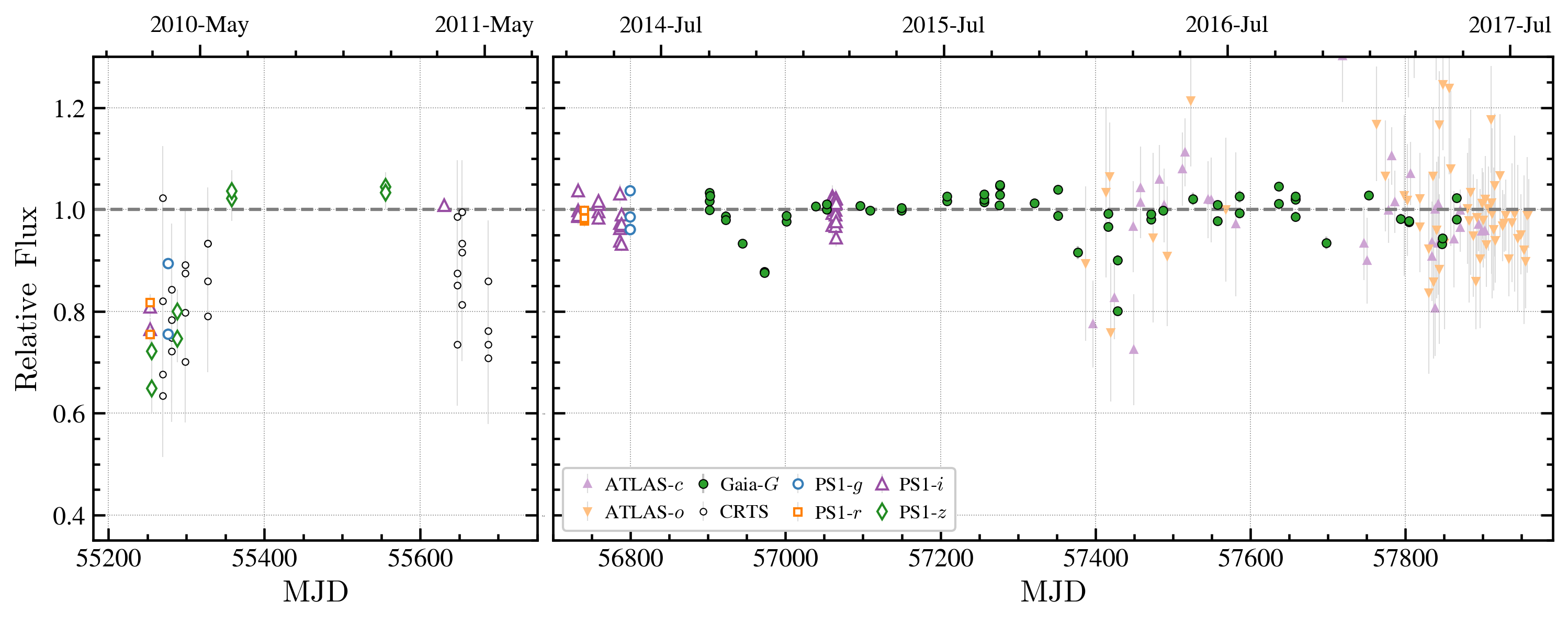}}
    \caption{Survey photometry for \tar\ centered around additional transits around 55300\,MJD (2010~April) detected in PanSTARRS (PS1) in the left panel and 57410\,MJD (2016~January) detected in Gaia in the right panel. All points have uncertainties; some error bars are smaller than the points themselves.}%
    \label{fig:zooms}
\end{figure*}

TESS data collected out-of-transit (2019, 2020, and 2022) do not show significant variability, but TESS data in 2024 collected just after the large dimming in event from 2023 reveal several significant periodicities. We have yet to confidently identify the physical origin of the periodic signals, but they appear intrinsic to \tar.

It is possible a surface inhomogeneity (spot) is responsible for the longest TESS photometric period at 14.842-hr; the shorter signals at 95.8416\,min and 8.57\,min would be extremely fast for a white dwarf rotation period \citep{2017ApJS..232...23H}. However, spotted white dwarfs tend to be fairly sinusoidal, rarely revealing higher than the third harmonic of the rotation period (e.g., \citealt{2017ApJ...835..277H,2020ApJ...897L..31W}). Sector~75 shows at least five harmonics of the 14.842-hr fundamental signal. The morphology of the periodogram is far more reminiscent of the many harmonics seen trying to reproduce the non-sinusoidal shape of transiting debris \citep{2021ApJ...917...41V,2022MNRAS.511.1647F}.

Should the 14.842\,hr signal we detect in TESS Sector~75 be the orbital period of the debris transiting \tar, it would be the fifth measured orbital period of transiting debris at a white dwarf: WD\,1145+017 -- 4.49\,hr \citep{2015Natur.526..546V}, ZTF\,J0328$-$1218 -- 9.937\,hr \& 11.2\,hr \citep{2021ApJ...917...41V}, WD\,1054$-$226 -- 25.02\,hr \citep{2022MNRAS.511.1647F}, ZTF\,J0139+5245 -- $\approx$\,107\,d \citep{2020ApJ...897..171V}. Altogether, this spectrum of periods may be evidence for the gradual evolution of tidally disrupted material from highly eccentric, extended, long-period ($\gtrsim$\,$10^2$\,d) orbits to closer-in, more circular orbits with periods comparable to the circular orbit at the tidal disruption radius for a typical white dwarf (e.g., \citealt{2022MNRAS.509.2404B}), $P_{\rm orbit, TD} \sim 4.5$\,hr.

In addition to the dramatic 2023 dimming event, Figure~\ref{fig:zooms} shows that \tar\ reveals further extended ($>$3\,month) transits from PanSTARRS data, deepest around 2010~April, as well as a transit in Gaia DR3, deepest around 2016~January. There does not appear to be a coherent recurrence timescale to these extended events, which recur roughly every $\sim$6-8\,years. Fixed patterns in the circumstellar gas in some white dwarf debris disks are measured to precess on years-long timescales \citep{2016MNRAS.455.4467M,2018ApJ...852L..22C}, despite also showing shorter-period signals (e.g., \citealt{2019Sci...364...66M}). However, the lack of coherence to the deep transits complicates precession as the core driver of the long-period recurrence of transits towards \tar.

We attempted to constrain the grayness of the debris transiting \tar. We emulate the analysis of \citet{2022AJ....163..263H} to search for trends in the $g-r$ colors among observations taken within 1-day intervals in ZTF. We fit two lines to the $g$ versus $g-r$ values: one with zero slope that would support a colorless, gray transit, and one where we allow the slope to vary. We find the reduced $\chi^2$ between the flat and sloped lines to be statistically consistent, with slight preference to the sloped line ($\chi^2_{\rm sloped} = 2.5$, $\chi^2_{\rm flat} = 3.2$). We further search for a trend between the ZTF-$g$ and ATLAS-$o$ photometry also within 1-day intervals, aiming to better resolve the 2023 transit with greater color contrast than can be achieved with $c-o$. Again, the reduced $\chi^2$ of the sloped line is preferred to the flat line, but are still comparable in value ($\chi^2_{\rm sloped} = 8.3$, $\chi^2_{\rm flat} = 11.3$). It does not appear there is a significant color term to the transits.

More stringently, the SDSS $u-r$, $u-i$, $g-r$, and $g-i$ colors are all consistent to within their associated uncertainties across the three SDSS photometric epochs. All bands were observed within 5\,min at each epoch. Figure~\ref{fig:allsurvey} shows all three epochs at different transit depths, one 5\% deep and one nearly 30\% deep. The standard deviation of the color term for the three epochs is smallest for $g-r$ and is 0.024\,mag, in line with the average $g-r$ photometric uncertainty of 0.027\,mag. The SDSS test suggests that the material is mostly gray, in line with other white dwarfs harboring transiting debris \citep{2016A&A...589L...6A,2018MNRAS.481..703I}.

We do not detect significant variability of the infrared excess from the warm, dusty debris disk around \tar, although its faintness would require very large ($>$20\%) changes to manifest as variable from the NEOWISE monitoring \citep{2024ApJ...972..126G}. NEOWISE observed \tar\ on 2023~May~4, which showed a 36\% and 26\% decrease in flux in $W1$ and $W2$, respectively, relative to the mean flux weighted to the photometric uncertainties. However, the large uncertainties means that these points only deviate 2.5$\sigma$ in $W1$ and 1.9$\sigma$ in $W2$ from their respective averages. An infrared dimming in transit would be consistent with gray transits. It could also suggest the extended events are unlikely collisionally driven, as they would generate a commensurate infrared brightening (e.g., \citealt{2021MNRAS.506..432S}).

Spectroscopy collected at various times and transit depths do not show significant abundance changes in the rock-forming elements (Ca, Mg, Fe, and Si) nor volatile elements (O and H) during various transit depths, although our in-transit measurements have the worst signal-to-noise and fewest species confidently detected. Triggering higher-resolution spectroscopy during a future deep transit could put much more stringent limits on the lack of in-transit spectroscopic variability in \tar, motivating monitoring of the system to see if there is optically thick circumstellar gas between us and the white dwarf (akin to the changing absorption features observed by \citealt{2020ApJ...897..171V} towards the long-period transiting white dwarf ZTF\,J0139+5245).

The 25-year light curve of \tar\ in Figure~\ref{fig:allsurvey} reveals that this transiting white dwarf can go for many years (perhaps half a decade) at quiescent flux levels before entering deep transit. As also noted by \citet{2024MNRAS.530..117A}, this long-term variability complicates accurate calculations of transit occurrence rates \citep{2018MNRAS.474.4603V,2019MNRAS.486.4574R,2024MNRAS.533.1756R}. This motivates extending the baseline for surveys like ZTF and ATLAS to keep watch for more transit events, and we eagerly await new facilities such as Rubin \citep{2019ApJ...873..111I} and Roman \citep{2019arXiv190205569A} to open this search for extended transits to orders of magnitude more faint white dwarf stars.

\section*{Acknowledgements}

We thank the anonymous referee for constructive feedback that helped this work, Tim~Cunningham for helpful discussions in the preparation of this manuscript, and J.~W.~Kuehne for observation support. This material is based upon work supported by the National Aeronautics and Space Administration under Grant No. 80NSSC23K1068 issued through the Science Mission Directorate. Support for this work was in part provided by NASA TESS Cycle 6 grant 80NSSC24K0878. J.\,A.\,G. is supported by the National Science Foundation Graduate Research Fellowship Program under Grant No. 2234657. 

The ZTF forced-photometry service was funded under the Heising-Simons Foundation grant \#12540303 (PI: Graham). This work has made use of data from the European Space Agency (ESA) mission Gaia (\url{https://www.cosmos.esa.int/gaia}), processed by the Gaia Data Processing and Analysis Consortium (DPAC, \url{https://www.cosmos.esa.int/web/gaia/dpac/consortium}). Funding for the DPAC
has been provided by national institutions, in particular the institutions participating in the Gaia Multilateral Agreement. This paper includes data collected by the TESS mission. Funding for the TESS mission is provided by the NASA's Science Mission Directorate. The CSS survey is funded by the National Aeronautics and Space Administration under Grant No. NNG05GF22G issued through the Science Mission Directorate Near-Earth Objects Observations Program. The CRTS survey is supported by the U.S.~National Science Foundation under grants AST-0909182. Some of the data presented herein were obtained at Keck Observatory, which is a private 501(c)3 non-profit organization operated as a scientific partnership among the California Institute of Technology, the University of California, and the National Aeronautics and Space Administration. The Observatory was made possible by the generous financial support of the W. M. Keck Foundation. The authors wish to recognize and acknowledge the very significant cultural role and reverence that the summit of Maunakea has always had within the Native Hawaiian community. We are most fortunate to have the opportunity to conduct observations from this mountain. 

S.\,X. is supported by NOIRLab, which is managed by the Association of Universities for Research in Astronomy (AURA) under a cooperative agreement with the National Science Foundation. M.\,L.\,K. and K.\,H. acknowledge support from the Wootton Center for Astrophysical Plasma Properties, a U.S. Department of Energy NNSA Stewardship Science Academic Alliance Center of Excellence supported under award numbers DE-NA0003843 and DE-NA0004149, from the United States Department of Energy under grant DE-SC0010623. K.\,H. acknowledges support from the National Science Foundation grant AST-2108736. This research was improved by discussions at the KITP Program ``White Dwarfs as Probes of the Evolution of Planets, Stars, the Milky Way and the Expanding Universe'' supported by National Science Foundation under grant No. NSF PHY-1748958.

\vspace{5mm}
\facilities{Zwicky Transient Facility (ZTF), Asteroid Terrestrial-impact Last Alert System (ATLAS), Gaia Collaboration, Pan-STARRS (PS1), Catalina Real-Time Transient Survey (CRTS), Sloan Digital Sky Survey (SDSS), Las Cumbres Observatory Global Telescope (LCOGT), Wide-field Infrared Survey Explorer (WISE), McDonald Observatory 2.1-Meter Otto Struve (ProEM), Transiting Exoplanet Survey Satellite (TESS), Hobby-Eberly Telescope Dark Energy Experiment (HETDEX), Palomar 200-inch Telescope (DBSP), Keck:I (LRIS)}

% Feel free to include any software used.

\software{astrobase \citep{wbhatti_astrobase}, astropy \citep{2013A&A...558A..33A,2018AJ....156..123A}, astroquery \citep{2019AJ....157...98G}, cecilia \citep{2024MNRAS.529.1688B}, lmfit \citep{newville_matthew_2014_11813}, Matplotlib \citep{Hunter:2007}, NumPy \citep{harris2020array}, pandas \citep{pandas2020}, SciPy \citep{2020SciPy-NMeth}, Period04 \citep{2005CoAst.146...53L}, phot2lc \citep{phot2lc}, TESS\_localize \citep{2023AJ....165..141H}}

%\bibliography{SBSS1232+563}{}
%\bibliographystyle{aasjournal}

\end{CJK}
\end{document}